\documentclass[a4paper]{article}

\usepackage[english]{babel}
\usepackage[utf8x]{inputenc}
\usepackage[T1]{fontenc}
\usepackage{amsmath,amssymb}
\usepackage{graphicx}
\usepackage[colorinlistoftodos]{todonotes}
\usepackage[colorlinks=true, allcolors=blue]{hyperref}
\usepackage{booktabs}

\newcommand{\ratio}{f_n/f_p}
\newcommand{\mdm}{M_{DM}}

\title{\bf \LARGE Isospin-violating dark matter in the light of recent data}
\author{Carlos E. Yaguna\\[3mm]
\textit{Max-Planck-Institut f\"ur Kernphysik}\\ \textit{Saupfercheckweg 1, 69117 Heidelberg, Germany}}
\date{}

\begin{document}
\maketitle

\begin{abstract}
In scenarios where the dark matter interacts differently with protons and neutrons (isospin-violating dark matter), the interpretation of the experimental limits on the dark matter spin-independent cross section may be significantly modified. On the one hand, the  direct detection constraints  are shifted depending on the target nucleus, possibly changing the hierarchy among  different  experiments.  On the other hand,  the relative strength between the bounds from neutrino detectors and those from direct detection experiments is altered, allowing the former to be more competitive. In this paper,  the  status of isospin-violating dark matter is assessed  in the light of recent  data, and the prospects for its detection in the near future are analyzed. We find, for example, that there are regions in the parameter space where   IceCube currently  provides the most stringent limits on the  spin-independent cross section, or others where the expected sensitivity of DEAP-3600 is well above the  LUX exclusion limit.  Our results highlight the complementarity among  different targets in direct  detection experiments,  and between direct detection and neutrino searches  in the quest for a  dark matter signal. 
\end{abstract}

\section{Introduction}

So far,  all the evidence  for the existence of dark matter comes exclusively from its  gravitational effects on baryonic matter \cite{Bertone:2004pz, Ade:2015xuaf, Queiroz:2016awc}. As a result, not much is known about the fundamental properties of the dark matter particle. Determining these properties --the dark matter nature-- is one of the most important open problems in particle and astroparticle physics today,  and there is a very active experimental program aimed precisely at it \cite{Buckley:2013bha,Undagoitia:2015gya}. 

In the last two years, for example,  new limits on the dark matter spin-independent cross section have been reported both by direct detection experiments \cite{Aprile:2016swn,Akerib:2016vxi,Tan:2016zwf,Akerib:2015rjg,Amole:2015pla,Agnes:2015ftt,Agnese:2015ywx} and by neutrino detectors \cite{icpre,Adrian-Martinez:2016gti,Aartsen:2016exj,Choi:2015ara} --the former try to directly detect the elastic scattering of dark matter particles off nuclei, whereas the latter search for a neutrino signal from the annihilation of dark matter particles captured by the Sun. Moreover, significant improvements to these limits  are expected  within the next few years from a new generation of direct detection experiments that are currently running \cite{Aprile:2015uzo,Fatemighomi:2016ree}. Usually, these experimental results are interpreted \emph{assuming} that the dark matter particle interacts in the same way with protons and neutrons, a hypothesis that does  not necessarily holds. 

Isospin-violating dark matter denotes a generic scenario where the dark matter coupling to the protons is different than that to  neutrons \cite{Kurylov:2003ra, Giuliani:2005my,Chang:2010yk, Kang:2010mh,Feng:2011vu}. It was initially proposed to understand the puzzling results  reported by some  direct detection experiments at low dark matter masses, but it was soon realized to be a more general framework that  provides a rich phenomenology also at higher masses \cite{Gao:2011bq}, $\mdm\gtrsim 30$ GeV. This high mass region is  the focus of this paper.  If the dark matter particle is  isospin-violating,  the interpretation of the experimental limits on the dark matter spin-independent cross section may be  modified in two important ways. First, the  direct detection constraints  are shifted depending on the target nucleus, possibly changing the hierarchy among  different  experimental limits.  Second,  the relative strength between the bounds from direct detection experiments and those from neutrino detectors  is altered, allowing the latter to be more competitive. 

In this paper,  an updated analysis of the status of isospin-violating dark matter  and of its detection prospects in the near future is presented. We incorporate into this study  the several experimental limits on the spin-independent cross section obtained in the last two years --from CDMS II, SuperKamiokande, DarkSide-50, PandaX, ANTARES, XENON100, IceCube and LUX-- as well as the projected sensitivity of currently operating direct detection experiments --DarkSide-50, DEAP-3600, and XENON1T. Comparing the resulting  limits and sensitivities for several values of the isospin-violating parameter,  we find significant differences with respect  to the expectations of the standard scenario. Above all, our results highlight  the complementarity among  different targets in direct  detection experiments,  and between direct detection and neutrino searches  in the quest for a  dark matter signal.

The rest of the paper is organized as follows. In the next section, the recent experimental limits on  the dark matter spin-independent cross section are described. Then, in section \ref{sec:iso}, we outline the isospin-violating scenario, emphasizing how the interpretation of experimental results is modified within this framework. Our main results are described in section \ref{sec:current} and are summarized in figures \ref{fig:newfnfp07}-\ref{fig:newfnfp08}. Finally, we draw our conclusions in section \ref{sec:con}. 
\section{Recent experimental results}
Here we  review the most important limits on  the dark matter-proton spin-independent cross section obtained recently (last two years or so), and their interpretation within the standard scenario, which assumes that the dark matter couples equally to proton and neutrons. For clarity, we will list them in chronological order, first for direct detection experiments and then for neutrino detectors. The reader who is already familiar with these developments can look at  figures \ref{fig:standardA} and \ref{fig:standardB}, where they are summarized, and jump directly to the next section. 

\subsection{Direct detection experiments}

\begin{figure}[tb]
\centering
\includegraphics[scale=0.7]{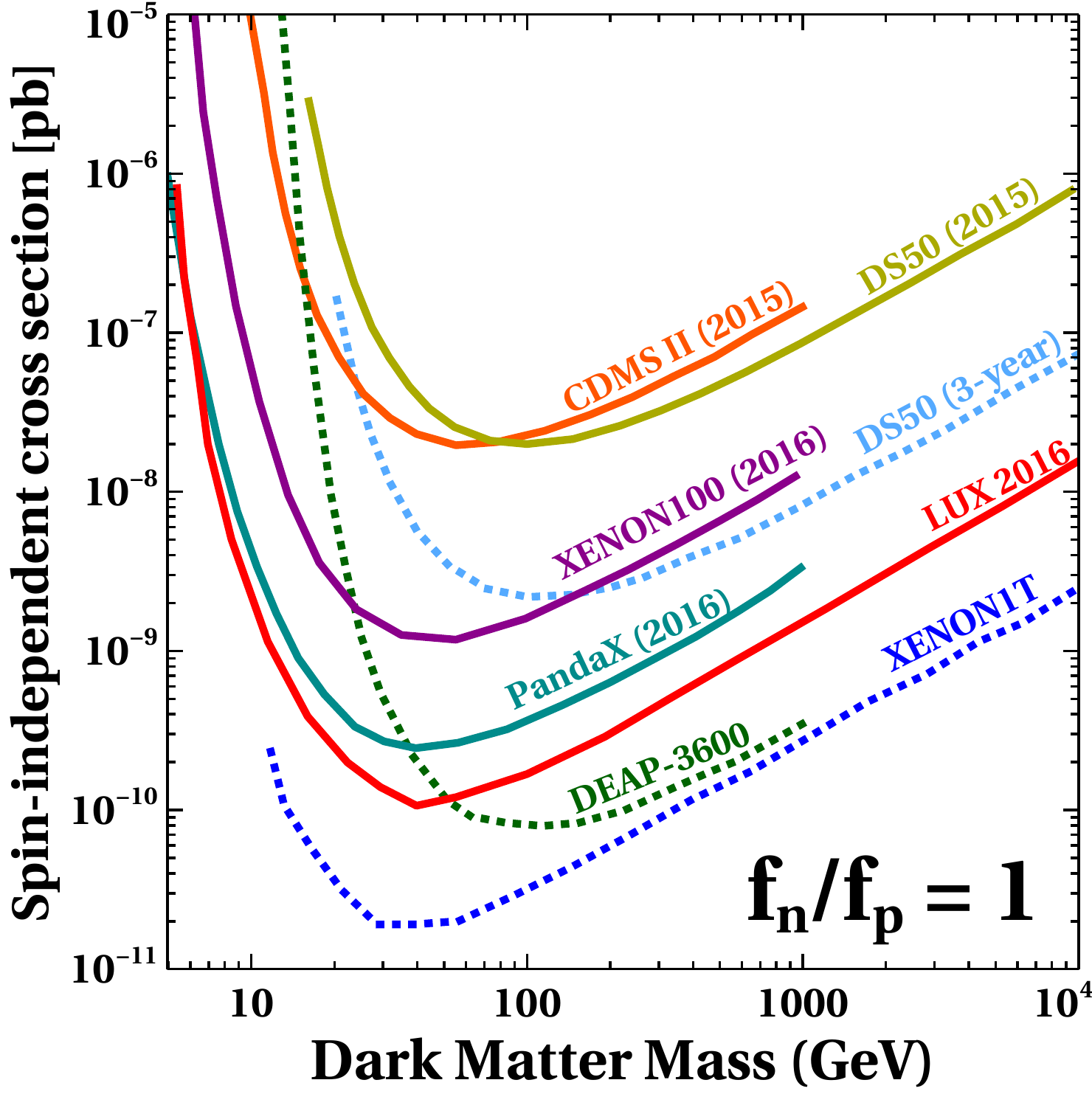}
\caption{The solid lines show the recent constraints on the dark matter-proton spin-independent cross section for $\ratio = 1$ (isospin-conservation) from different direct detection experiments: CDMS II (orange), DS50 (yellow), XENON100 (magenta), PandaX (cyan), and LUX (red). The dotted lines show instead the expected sensitivity in currently running direct detection experiments: DS50 (light blue), DEAP-3600 (green), and XENON1T (blue). \label{fig:standardA}}
\end{figure}

Let us take as our starting point the beginning of 2015. At that time, the most stringent limits on the dark matter spin-independent cross section were the first results of the LUX  experiment \cite{Akerib:2013tjd}, reported  in October 2013. They improved the  XENON100 limits from July 2012 \cite{Aprile:2012nq} by about a factor two, and  were significantly stronger than the limits from all the other experiments.

In April 2015 the superCDMS collaboration presented a new analysis \cite{Agnese:2015ywx} of the data from the Ge detectors in the final CDMS II five-tower exposure, with  a total raw exposure of about 612 kg days. This data had been acquired between July 2007 and September 2008 and had set the world-leading sensitivity to the spin-independent direct detection cross section in 2010, when it was first analyzed. Thanks to the new data reduction algorithms and surface-event rejection methods, the new analysis improved that limit  by more than a factor 2 at high dark matter masses.  In figure \ref{fig:standardA} this new  CDMS II limit is shown as a solid orange line.

In October 2015, the DarkSide collaboration reported the first limits on the WIMP-nucleon spin-independent cross section, obtained by the DarkSide-50 dark matter detector, using a target of low-radioactivity (underground) argon \cite{Agnes:2015ftt}. This search was based on data accumulated over 70.9 live-days and improved the bound based on atmospheric argon published by the same collaboration a year earlier \cite{Agnes:2014bvk}. When the null results of both searches, with atmospheric and underground argon, are combined, the limit denoted in figure \ref{fig:standardA} as DS50 (2015) is obtained.

That  same  month, the dark matter search results from the PICO-60 detector, a $\mathrm{CF_3I}$ bubble chamber,  were published \cite{Amole:2015pla}. Regarding spin-independent interactions, the limits obtained by the PICO collaboration are weaker than those from CDMS II over the entire range of dark matter masses, and weaker than those from DarkSide-50 for dark matter masses larger than about 40 GeV.

In December 2015, a reanalysis of the 2013 data \cite{Akerib:2013tjd}, including $1.4\times 10^4$ kg day of search exposure, was issued by the LUX collaboration \cite{Akerib:2015rjg}. Thanks to the several advances incorporated into this new analysis, a 23\% improvement in sensitivity at high WIMP masses over the first LUX result (published two years earlier) was achieved.  These results provided  the most stringent limits on the spin-independent WIMP-nucleon cross section at that time and were superseded by new LUX limits published seven months later.

In February 2016, the results of the commissioning run of the PandaX-II experiment, a half-ton scale dual-phase xenon experiment located at the China JinPing underground Laboratory (CJPL), were presented \cite{Tan:2016diz}. This  search was based on data taken between November and December 2015, with a total exposure of $306\times 19.1$ kg day. The constraints obtained on the spin-independent cross section were weaker, over the entire range of dark matter masses, than those  previously set by the LUX experiment and were, in any case, updated soon afterwards. 

On July 21st 2016, the LUX collaboration released, at the IDM 2016 conference in Sheffield, UK,  the results from the detector's final run \cite{luxidm}. These results were based on data collected  from October 2014 to May 2016, with a total exposure of $3.35\times 10^4$ kg day (332 live days).  Regarding spin-independent limits, they  yielded  about a  factor $4$ improvement in sensitivity (at high dark matter masses) with respect to the LUX results from December 2015, which  then provided the most stringent constraints.  

Four days later, the PandaX-II experiment published  new limits on WIMP searches \cite{Tan:2016zwf}. This new  analysis was based on a total exposure of $3.3\times 10^4$ kg day and the resulting limits were very similar to those   announced by the LUX collaboration at IDM 2016. This limit is shown as a solid dark cyan line in figure \ref{fig:standardA}.

At the end of  August 2016, the LUX collaboration published the results that had been released  the month before, and presented new limits based on a combination of the new data with the previous LUX  exposure (data from 2013) \cite{Akerib:2016vxi}.  This combined limit is  more stringent than that from the  PandaX-II experiment  for all  values of the dark matter mass and it currently sets the most stringent constraints on the spin-independent cross section.  In figure \ref{fig:standardA}, this limit is displayed as a solid red line.

Lastly, in September 2016, the final results from the XENON100 experiment were reported \cite{Aprile:2016swn}.  They were  based on data taken between January 2010 and January 2014, with a total exposure of 477 live days (48 kg year).  Compared to the previous XENON100 results, this new analysis improves the limits on the spin-independent cross section  by about a factor $2$, and is shown as a solid dark magenta line in figure \ref{fig:standardA}. It is, however, significantly weaker than the constraints already set by the PandaX and LUX experiments. 

Figure \ref{fig:standardA} summarizes the most important recent limits (for our analysis) on the dark matter-proton spin-independent cross section. It can be seen there  that, when the dark matter interacts with protons and neutrons in the same way (isospin-conserving or $\ratio=1$), the current limits are largely dominated by experiments using Xenon as the target nucleus --LUX, PandaX, XENON100. The limits from experiments using Argon (DS50) or Germanium (CDMS II) are between one and two orders of magnitude weaker.  One of the goals of this paper is to examine whether this conclusion  still holds in the case of isospin-violating dark matter.

For comparison, in figure \ref{fig:standardA}, the expected sensitivity of currently running direct detection experiments are also displayed as dotted lines: DS50 (ligth blue) \cite{ds50-3year}, DEAP-3600 (green) \cite{Fatemighomi:2016ree}, and XENON1T (blue) \cite{Aprile:2015uzo}. Notice that DEAP-3600, an Argon detector, and XENON1T have similar expected sensitivities at high dark matter masses, a result that could  easily changed in the presence of isospin-violating interactions.

\subsection{Neutrino detectors}

\begin{figure}[tb]
\centering
\includegraphics[scale=0.7]{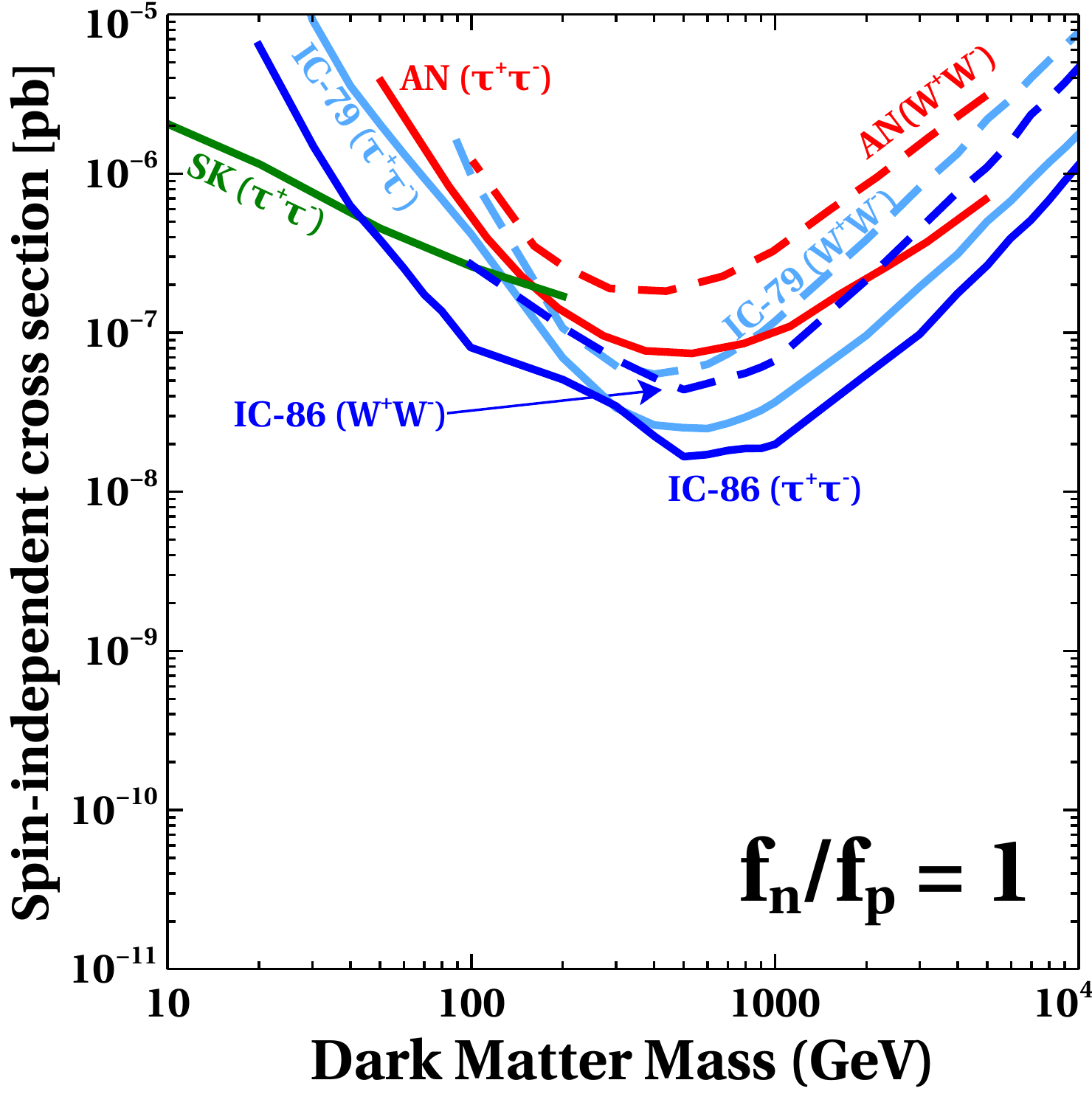}
\caption{\label{fig:standardB} A summary of the constraints on the dark matter-proton spin-independent cross section  obtained recently in neutrino detectors.  The colors of the lines distinguish the experiments --Super-Kamiokande (green), ANTARES (red), and IceCube (light and dark blue)-- whereas the type of line indicates the final state from dark matter annihilations: $\tau^+\tau^-$ (solid), and $W^+W^-$ (dashed). To facilitate the comparison with figure \ref{fig:standardA}, we have kept the same scale.}
\end{figure}

Neutrino detectors can constrain the dark matter spin-independent direct detection cross section by  searching for the neutrinos produced by dark matter annihilations inside the Sun \cite{Srednicki:1986vj}. When equilibrium between the WIMP capture and annihilation processes in the Sun is reached, which is often the case, the neutrino signal is determined only by the annihilation final states and by the capture rate, which is proportional to the direct detection cross section. For definiteness, we consider only the annihilation channels $W^+W^-$ and $\tau^+\tau^-$, which produce a \emph{hard} neutrino spectrum. Next, the most recent limits are reviewed.

In March 2015, the Super-Kamiokande collaboration published  new dark matter limits  based on 3903 days of SK data \cite{Choi:2015ara}.  These constraints on the spin-independent cross section are shown, for the $\tau^+\tau^-$ final state, as a green solid line in figure \ref{fig:standardB}, and  apply only for dark matter masses below 200 GeV.     

That same month, the IceCube collaboration presented \cite{Rameez:2015pph}, at the 50th Rencontres de Moriond, preliminary results based on data collected between May 2011 and May 2012 in the completed 86-string configuration of IceCube-DeepCore (341 days of livetime).  These new results improved the previous IceCube limits \cite{Aartsen:2012kia} by a factor between $30\%$ and $60\%$.  

In January 2016, the IceCube collaboration reported updated limits on dark matter annihilation in the Sun \cite{Aartsen:2016exj}. They reanalysed the data from the 79-string IceCube search that had been initially published in 2012 \cite{Aartsen:2012kia}, improving the analysis in several ways. In the paper, they considered different annihilation channels but presented results only for spin-dependent interactions, where IceCube is quite competitive with direct detection experiments.  To obtain the corresponding limits for the spin-independent interactions, we have employed the nulike code  provided in the same paper, which relies on  the public IC79 event information and detector response, and implements the same likelihood analysis used in the official paper \cite{Scott:2012mq}. The limits thus obtained  are shown in figure \ref{fig:standardB} as light blue lines  (denoted as IC-79) for annihilation into  $\tau^+\tau^-$ (solid line) and $W^+W^-$ (dashed line).

In March 2016, the results of a new analysis searching for dark matter annihilation in the Sun was reported by the ANTARES collaboration \cite{Adrian-Martinez:2016gti}. It was based on ANTARES data taken from 2007 to 2012 and the results were interpreted in terms of both spin-dependent and spin-independent interactions. The limits on the spin-independent interactions are displayed in figure \ref{fig:standardB} as red lines and are denoted by $\mathrm{AN(\tau^+\tau^-)}$  and $\mathrm{AN(W^+W^-)}$ for the two annihilation final states we consider. Notice that the ANTARES limits are not as stringent as the ones we derived based on the  IC-79 data that had been published earlier.

Finally, in September 2016, preliminary results from the final IceCube configuration were presented at the TeV Particle Astrophysics conference \cite{icpre}. They are based on 3 years of data taking (532 days of livetime) with the full IceCube detector (86 strings). The resulting limits are shown in figure \ref{fig:standardB} as solid ($\tau^+\tau^-$) and dashed ($W^+W^-$) blue lines and denoted as IC-86.   

Figure \ref{fig:standardB} summarizes the current limits on the dark matter spin-independent cross section from neutrino experiments. As can be seen there, for dark matter masses above 40 GeV, these limits are dominated by  the IceCube bounds. To facilitate the comparison with the limits from direct detection experiments, we have kept the same scale as in figure \ref{fig:standardA}. From these two figures, it becomes clear that, for isospin-conserving dark matter ($\ratio=1$), the limits on the spin-independent cross section from direct detection experiments are significantly stronger than those from neutrino detectors. It remains to be seen whether that is still the case for isospin-violating dark matter. 
\section{Isospin-violating dark matter \label{sec:iso}}
By definition, isospin-violating dark matter is a generic framework that includes all dark matter candidates that interact differently with protons and neutrons. The first example of such a dark matter particle to have been studied was likely  the heavy Dirac neutrino (see e.g. \cite{Primack:1988zm}), which couples more strongly to the neutrons than to the protons \cite{Gondolo:1996qw}. Throughout the years, many other candidates of this type have been proposed and their phenomenology has been  investigated in different contexts. In this section we first revisit general aspects of isospin-violating dark matter and then go over how the experimental results should be reinterpreted within this scenario. Finally, our main results are obtained and discussed in section \ref{sec:current}.

\subsection{Generalities}
Early works on isospin-violating dark matter were mostly motivated by tantalizing signals from direct detection experiments at low dark matter masses \cite{Kurylov:2003ra, Giuliani:2005my,Chang:2010yk, Kang:2010mh}. And it was in that context that the term \emph{Isospin-violating dark matter} was first introduced, in \cite{Feng:2011vu}. It has been realized since then, though, that isospin-violating dark matter is a more generic framework \cite{Feng:2013vaa} with a very interesting phenomenology. In practice, this framework extends the standard one with just one additional parameter: the ratio between the dark matter couplings to the neutron and to the proton.

Several explicit models of isospin-violating dark matter have been proposed in recent years. For scalar dark matter, a model with  colored mediators was studied in \cite{Hamaguchi:2014pja}  while a realization within a two-Higgs doublet model was considered in \cite{Drozd:2015gda}. For Dirac dark matter,  a model with a light $Z^\prime$ was investigated in \cite{Frandsen:2011cg}, a double portal scenario was proposed in \cite{Belanger:2013tla}, and a string-theory inspired construction  was  analyzed in \cite{Martin-Lozano:2015vva}. In \cite{Kang:2010mh,Gao:2011ka,Feng:2013vaa}, different supersymmetric realizations of isospin-violating dark matter were put forward. And models of asymmetric dark matter, for scalar and fermion dark matter, were examined in \cite{Okada:2013cba}.

Different experimental constraints on these scenarios have also been discussed \cite{Kumar:2011dr,Jin:2012jn,Hagiwara:2012we,Feng:2013fyw}. The intriguing interplay between direct detection searches and neutrino signals from the Sun was first discussed in \cite{Chen:2011vda} and \cite{Gao:2011bq}. Our analysis  is, in fact, similar in spirit to that presented in \cite{Gao:2011bq}. The main difference being the availability of new experimental results, as emphasized in the previous section. 

We now turn precisely to the question of how these experimental results should be reinterpreted for isospin-violating dark matter.
\subsection{Interpretation of experimental results}
\begin{figure}[tb]
\centering
\includegraphics[scale=0.7]{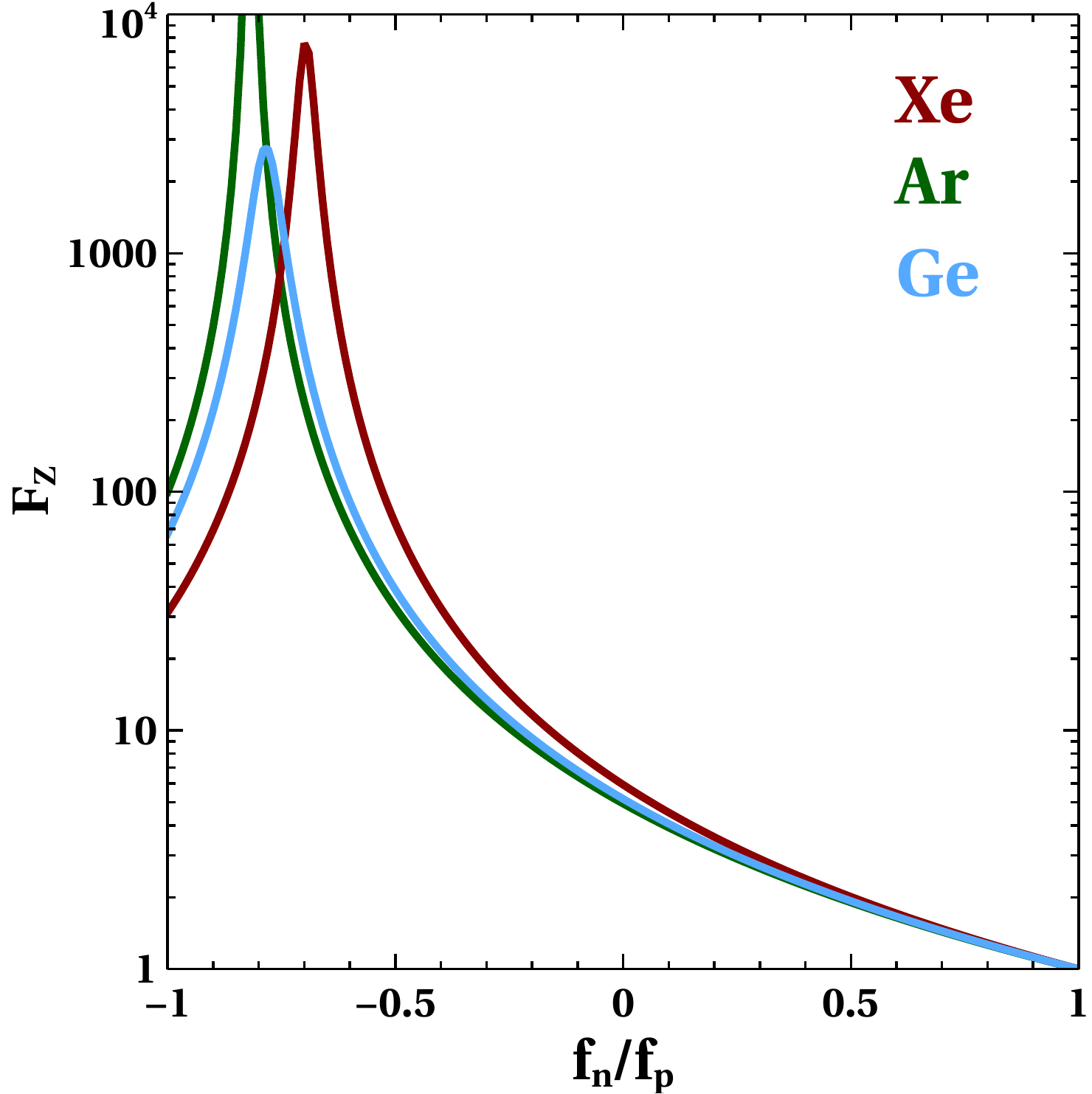}
\caption{\label{fig:ddxenon} The function $F_Z(\ratio)$ for  Xenon, Argon, and Germanium targets and $\ratio \in (-1,1)$.  $F_Z(\ratio)$ is the factor by which the sensitivity of a direct detection experiment to the dark matter-proton spin-independent cross section is reduced for isospin-violating dark matter.}
\end{figure}

The issue of how the experimental limits on the spin-independent cross section, from direct detection experiments and from neutrino detectors, should be modified for isospin-violating dark matter has been addressed before --e.g. in \cite{Gao:2011bq}. For completeness, we will summarize that discussion here. As usual for these scenarios, it is assumed in the following that the dark matter particle has only spin-independent interactions.

The event rate, $R$, at a dark matter detector can be written as
\begin{equation}
R= \sum_i \eta_i \sigma_{A_i}\, I_A
\end{equation}
where $A$ denotes the target nucleus and the sum is over the isotopes $A_i$ with fractional number abundances $\eta_i$. $I_A$ includes the astrophysical, experimental and nuclear physics inputs and is explicitly given by
\begin{equation}
I_A= n_{DM} N_T\int dE_R\int_{v_{min}}^{v_{max}}d^3v\, f(v) \frac{m_A}{2v\mu_A^2}F_A^2(E_R)
\end{equation}
where  $n_{DM}$ is the local  number density of DM particles, $N_T$ is the number of target nuclei, $f(v)$ is the DM velocity distribution, $v$ is the velocity of the DM particle,  and $F_A$ is the nuclear form factor. $\sigma_{A_i}$ is the dark matter-nucleus spin-independent cross section,
\begin{equation}
\label{eq:sigmaM}
\sigma_{A_i}=\frac{4\mu_{A_i}^2}{\pi}\left[f_{p}\,Z+f_{n}\,(A_i-Z)\right]^2\,,
\end{equation}
where $ \mu_{A_i}=\mdm M_{A_i}/(\mdm+M_{A_i})$ is the dark matter-nucleus reduced mass, $Z$ is the nucleus charge, and $A_i$ is the  number of nucleons. $f_p$ and $f_n$ denote respectively the couplings between the dark matter and the protons or neutrons, and are determined by the underlying particle physics model that accounts for the dark matter.  In particular, the dark matter-proton spin-independent cross section is given by
\begin{equation}
\label{eq:sigmap}
\sigma_p=\frac{4\mu_p^2}{\pi}f^2_{p}\,.
\end{equation}
Typically, direct detection experiments report their exclusion limits neither  in terms of $\sigma_{A_i}$ nor $\sigma_p$ but rather of $\sigma_N^Z$, the dark matter-nucleon cross section \emph{assuming} isospin conservation, which can be written as
\begin{equation}
\sigma_N^Z=\sigma_p\frac{\sum_i \eta_i\mu_{A_i}^2\left[Z+(A_i-Z)\ratio\right]^2}{\sum_i \eta_i \mu_{A_i}^2A_i^2}.
\end{equation}
Notice that  $\sigma_N^Z=\sigma_p$ for $f_n=f_p$ (isospin-conservation)  but in general this is not the case and one can have that $\sigma_N^Z\ll\sigma_p$. When the dark matter interactions violate isospin, it is $\sigma_p$ that is physically meaningful and that should be used to present the experimental results. It is convenient, therefore, to define the ratio
\begin{equation}
F_Z\equiv\frac{\sigma_p}{\sigma_N^Z},
\end{equation}
which is the factor by which the sensitivity of a direct detection experiment is suppressed when the dark matter interactions violate isospin. That is, if $\tilde\sigma$ is the limit (at a given $\mdm$) on the spin-independent  dark matter-nucleon cross section reported by an experiment,  then $F_Z \tilde\sigma$ is the corresponding limit on the dark matter-proton cross section that actually applies to the isospin-violating scenario (in some cases,  this simple picture may be  modified \cite{Cirigliano:2013zta,Zheng:2014nga}). $F_Z$ is a function that depends on the target nucleus and on the isospin-violating parameter $\ratio$. Figure \ref{fig:ddxenon} shows $F_Z$ for Xenon, Argon and Germanium targets, and values of $\ratio$ between $-1$ and $1$. Since Argon consist mostly of a single isotope, $F_Z$ goes to arbitrary high values at the Argophobic point, $\ratio=-0.82$, whereas for $\mathrm{Xe}$ and $\mathrm{Ge}$, which consist of several isotopes, $F_Z$ has a maximum (the relevance of the distribution of isotopes present in each detector was first emphasized in \cite{Feng:2011vu}). For a Xenon target, this maximum value of $F_Z$ is achieved  for $\ratio\approx -0.7$ (the so-called Xenophobic point) and amounts to  about $10^4$. Hence, the constraints on the spin-independent cross section from  Xenon experiments (XENON100, Panda-X, LUX) can be relaxed by up to four orders of magnitude with respect to the results presented by the collaborations.  From this figure it can be seen that the results of direct detection experiments can be substantially modified only for negative values of $\ratio$ and only within a narrow range of values. For that reason, in our analysis particular attention will be paid to the region  $\ratio\in (-0.6,-0.8)$, which includes the Xenophobic point.    

\begin{table}[tb]
\centering
\begin{tabular}{ccrrrr}
\toprule
$\mdm$ (GeV) & $\ratio=-0.80$  & $-0.75$ & $-0.70$ & $-0.65$ & $-0.60$  \\\midrule
10 & 43.5 & 35.9 & 29.4 & 24.1 & 19.9\\
20 & 65.0 & 49.5 & 38.0 & 29.7 & 23.6\\
30 & 76.2 & 55.6 & 41.5 & 31.7 & 24.8\\
40 & 82.8 & 58.9 & 43.2 & 32.7 & 25.4\\
50 & 87.1 & 61.0 & 44.2 & 33.2 & 25.7\\
60 & 90.2 & 62.3 & 44.9 & 33.5 & 25.9\\
70 & 92.4 & 63.3 & 45.3 & 33.8 & 26.0\\
80 & 94.2 & 64.0 & 45.7 & 33.9 & 26.1\\
90 & 95.6 & 64.6 & 45.9 & 34.1 & 26.2\\
100 & 96.7 & 65.1 & 46.1 & 34.1 & 26.2\\
200 & 101.8 & 66.9 & 46.9 & 34.5 & 26.3\\
1000 & 104.7 & 67.7 & 47.0 & 34.5 & 26.3\\
10000 & 105.1 & 67.7 & 47.0 & 34.4 & 26.2\\
\bottomrule
\end{tabular}
\caption{\label{tab:ntsup} The function $F_\odot(\mdm,\ratio)$ for selected  values of $\mdm$ and $\ratio$.  $F_\odot(\mdm,\ratio)$ is the factor by which the sensitivity of a neutrino telescope to the  dark matter-proton spin-independent cross section is reduced for isospin-violating dark matter.}
\end{table}

Let us now switch to neutrino detectors. They  search for a signal from the annihilation of the dark matter particles that have been captured by the Sun. When the capture rate $\Gamma_C$ and the annihilation rate ($\Gamma_A$) are in equilibrium,  the event rate depends on the dark matter properties only through $\Gamma_C$ and the annihilation final state.  Since $\Gamma_C$ is determined by the  dark matter spin-independent scatterings with the nuclei in the Sun, it can be parameterized by
\begin{equation}
\Gamma_C=\sigma_p\, C_0(\mdm,\ratio)
\end{equation}
where $C_0(\mdm,\ratio)$ is a capture coefficient. Given that  the limits from neutrino detectors on the dark matter spin-independent cross section are usually reported assuming isospin invariance (a notable exception being \cite{Choi:2015ara}),  it is useful to define, in analogy with $F_Z$, the function
\begin{equation}
F_\odot=\frac{C_0(\mdm,\ratio=1)}{C_0(\mdm,\ratio)},
\end{equation}
which is the factor by which the sensitivity of a neutrino detector is suppressed when the dark matter interactions do not conserve isospin. Hence, if $\tilde\sigma$ is the limit on the dark matter-nucleon cross section derived by a neutrino detector, $F_\odot \tilde\sigma$ is the actual limit on $\sigma_p$ that applies to the isospin-violating scenario.  Table \ref{tab:ntsup} displays the values of $F_\odot$ for selected values of the dark matter mass and of $\ratio$ (around $-0.7$).  To evaluate the capture coefficients in the Sun, $C_0(\mdm,\ratio)$, we relied on the model-independent routines implemented  in  micrOMEGAs \cite{Belanger:2013oya}, which take into account the contributions from all nuclei up to $\mathrm{^{59}Ni}$. Notice that for the  values  shown in table \ref{tab:ntsup}, $F_\odot$ amounts at most to about a factor $100$. That is, the constraints from neutrino detectors can be degraded by up to two orders of magnitude for isosping-violating interactions. 

By comparing figure \ref{fig:ddxenon} and table \ref{tab:ntsup}, one can see that there are regions in the parameter space of isospin-violating dark matter where the limits from direct detection experiments are weakened more than the limits from neutrino detectors, allowing the latter to be more competitive than in the standard scenario. 
\subsection{Current limits and future prospects \label{sec:current}}

\begin{figure}[tb]
\centering
\includegraphics[scale=0.7]{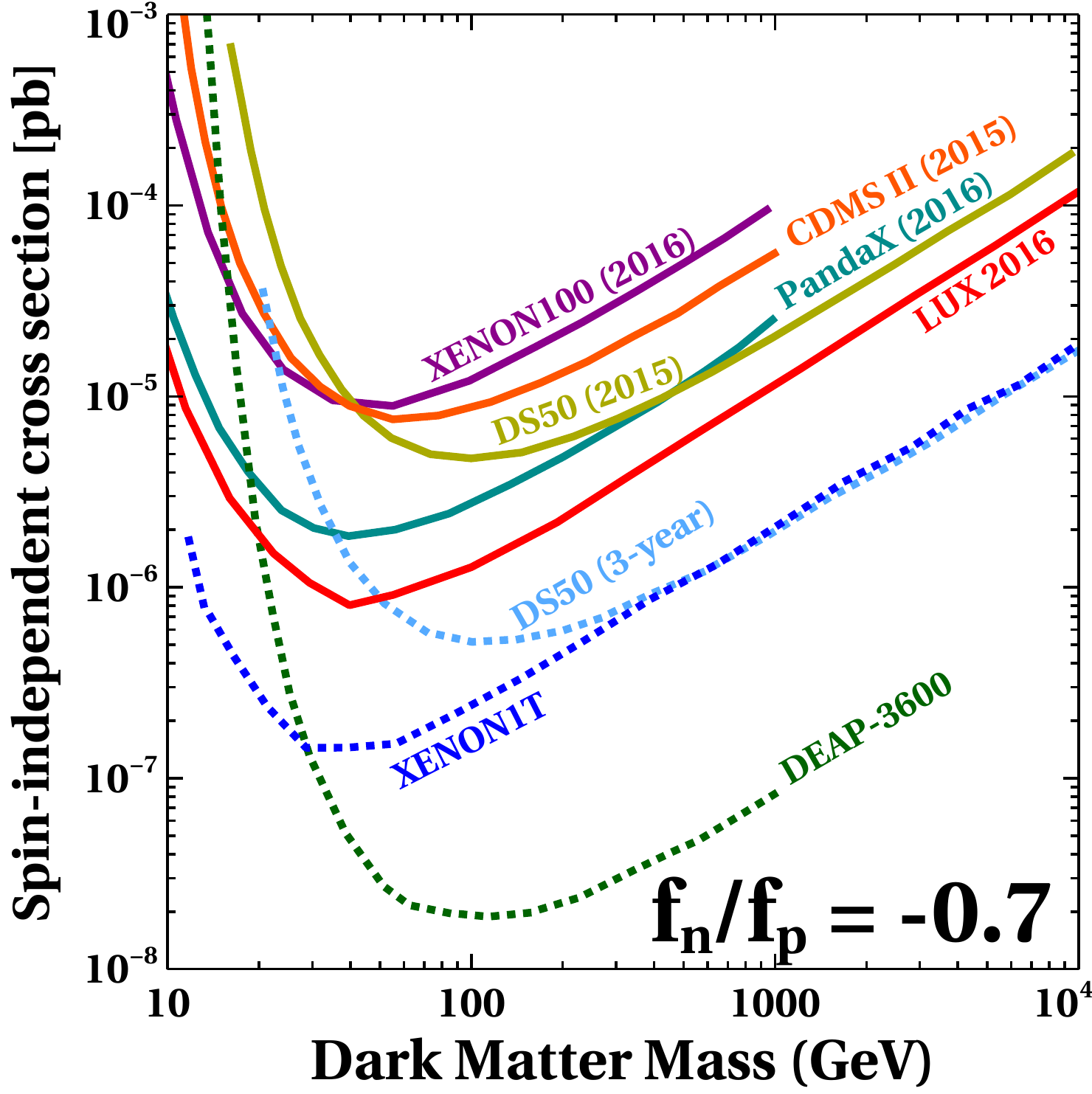}
\caption{\label{fig:newfnfp07} Current limits and expected sensitivities to the dark matter-proton spin-independent cross section for $\ratio = -0.7$. This value of $\ratio$  maximally suppresses the sensitivity of  Xenon targets.  The conventions are the same as in figure \ref{fig:standardA}. Notice that, for this value of $\ratio$, the current bound from DarkSide50 is comparable to the one from LUX at high masses. In addition, DEAP-3600 is expected to probe a much larger region of parameter space than XENON1T. }
\end{figure}

It is now time to put together what has been discussed so far in order to determine the current status of isospin-violating dark matter, and to investigate its detection prospects in the near future.   As we have already argued, significant modifications to the standard results are expected only within a narrow range of $\ratio$, so we will limit our analysis to $\ratio$ between $-0.6$ and $-0.8$.

To begin, let us consider the Xenophobic case first, $\ratio=-0.7$, for which the direct detection limits from Xenon experiments are maximally degraded --see figure \ref{fig:ddxenon}. It should be emphasized, though, that all direct detection limits will be relaxed to some extent. Figure \ref{fig:newfnfp07} summarizes the current limits and the expected sensitivities for this case. First of all, notice that the region of the parameter space that can be probed is much smaller, with current limits extending only down to $10^{-6}$ pb and future experiments reaching about $10^{-8}$ pb. Regarding the current status, from the figure we see that the CDMS II limit turns out to be stronger than the XENON100 limit for dark matter masses above $30$ GeV. Likewise, the DarkSide-50 limit becomes more stringent than the PandaX-II for dark matter masses larger than $400$ GeV, and, at high dark matter masses, it is weaker than the LUX constraint only by about a factor two. Interestingly, even in this case, where Xenon targets are maximally penalized, the most stringent limits on the spin-independent cross section are set by a Xenon experiment --an indication of how dominant Xenon detectors have become in direct detection searches.

Regarding detection prospects, the situation looks more interesting. On the one hand, the sensitivity of DarkSide-50 is expected to be comparable, for dark matter masses above 300 GeV, to that of the much larger XENON1T.  On the other hand,  DEAP-3600 will reach cross sections that are up to one order of magnitude beyond the expected sensitivity of XENON1T. Hence, if this value of $\ratio$ is realized in nature, it is the DEAP-3600 experiment that has the best prospects to detect a dark matter signal in the near future.

\begin{figure}[tb]
\centering
\includegraphics[scale=0.7]{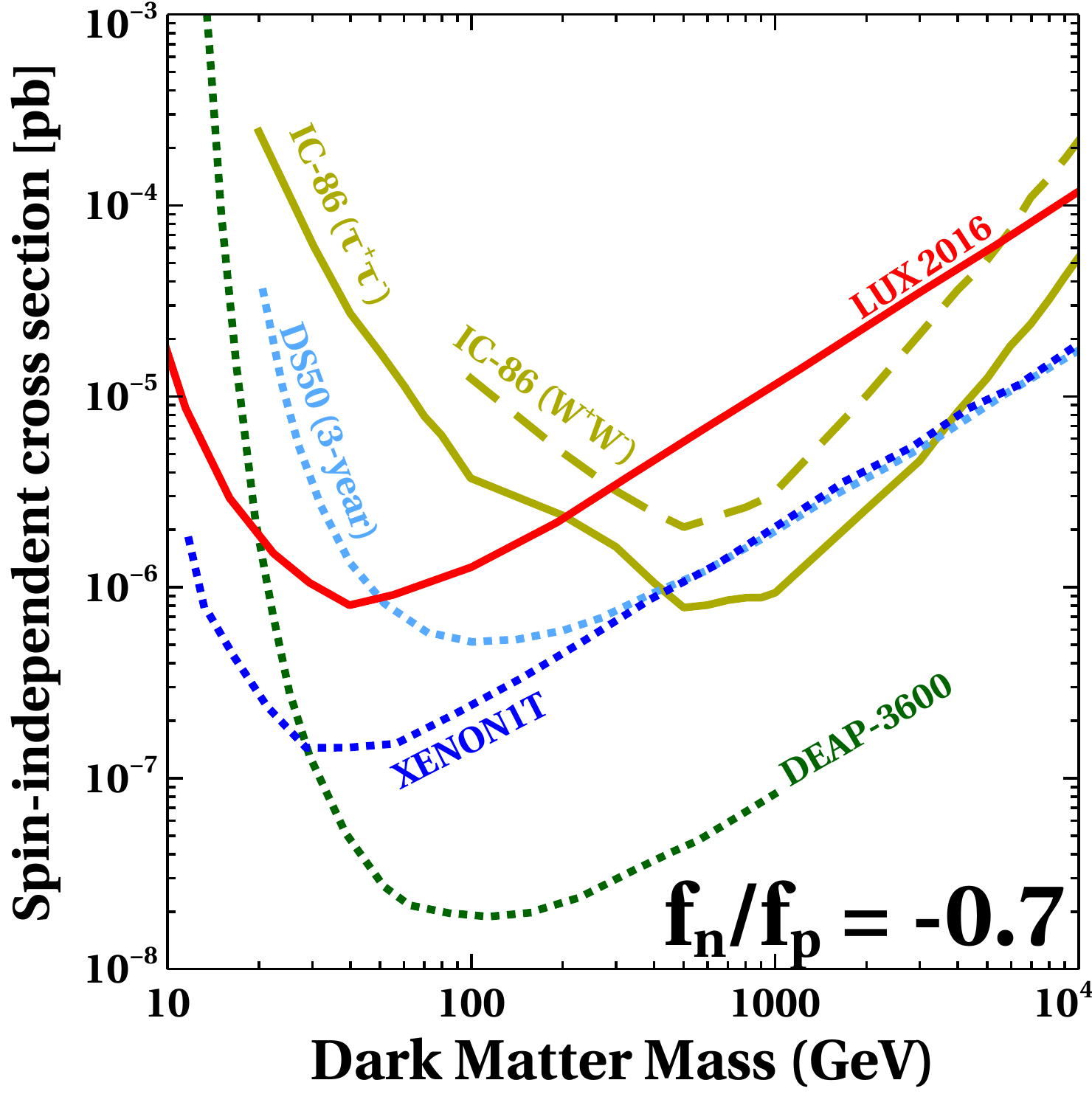}
\caption{\label{fig:fnfp07B} A comparison between direct detection and IceCube  bounds on the dark matter-proton spin-independent cross section for $\ratio = -0.7$. Notice that, in this case, IceCube bounds can be significantly stronger than the  direct detection ones. In fact, IceCube has already excluded some regions of the parameter space that lie beyond the expected reach of XENON1T.}
\end{figure}

Figure \ref{fig:fnfp07B} compares, for $\ratio = -0.7$, the best limit from direct detection experiments against the IceCube limits for the $\tau^+\tau^-$ and $W^+W^-$ final states.  Interestingly, we see that IceCube data is  probing some regions of the parameter space which are not excluded by direct detection experiments. In other words,  IceCube currently sets the most stringent limit,  over a wide range of dark matter masses, on the spin-independent cross section for dark matter annihilations into $\tau^+\tau^-$ and $W^+W^-$. This result is one of the most important findings of our analysis. Regarding future prospects, we see from the figure that a non-negligible fraction of the parameter space that could be probed by DarkSide-50 and XENON1T is already excluded by IceCube data. DEAP-3600 will instead be able to test significant regions of new parameter space not currently probed by other experiments. Notice in particular that for dark matter masses around 200 GeV and $\sigma_p\sim 10^{-6}$ pb, all three currently running direct detection experiments could find a dark matter signal. 
 
\begin{figure}[tb]
\centering
\includegraphics[scale=0.6]{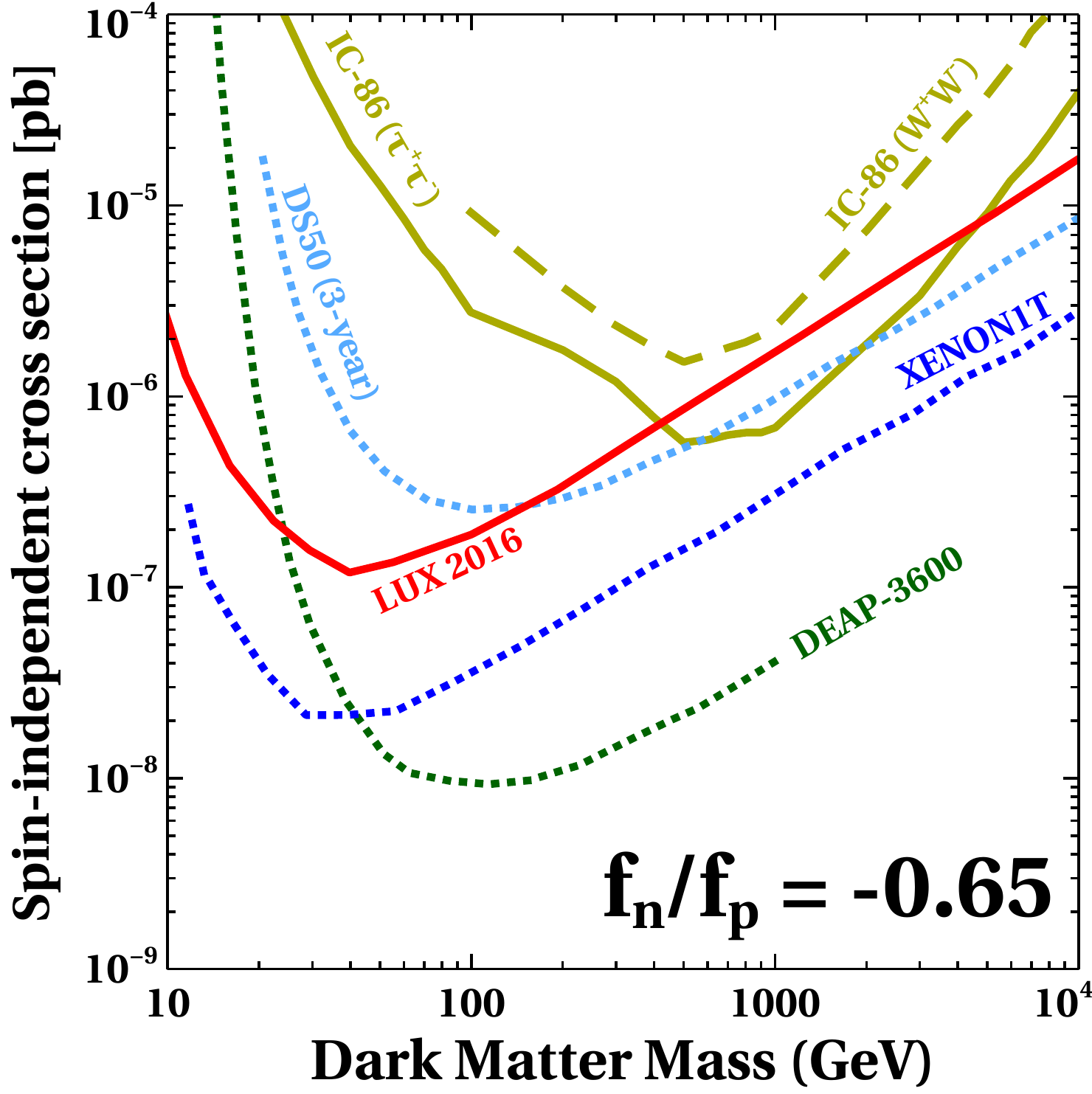} 
\caption{\label{fig:newfnfp065} A comparison between direct detection and IceCube  bounds on the dark matter-proton spin-independent cross section for $\ratio = -0.65$}
\end{figure}

\begin{figure}[tb]
\centering
\includegraphics[scale=0.6]{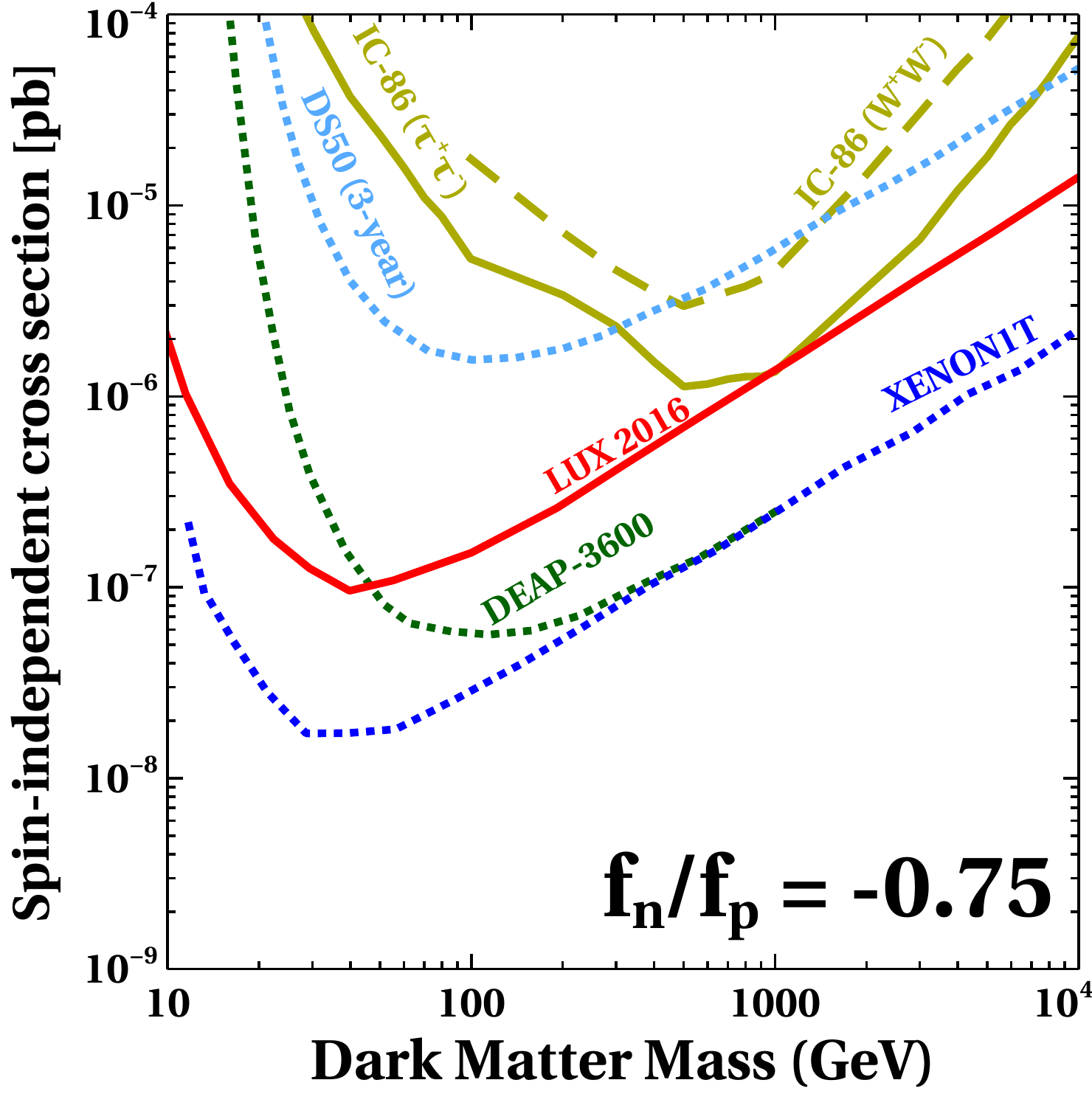}
\caption{\label{fig:newfnfp075} A comparison between direct detection and IceCube  bounds on the dark matter-proton spin-independent cross section for  $\ratio = -0.75$.}
\end{figure}

\begin{figure}[tb]
\centering
\includegraphics[scale=0.6]{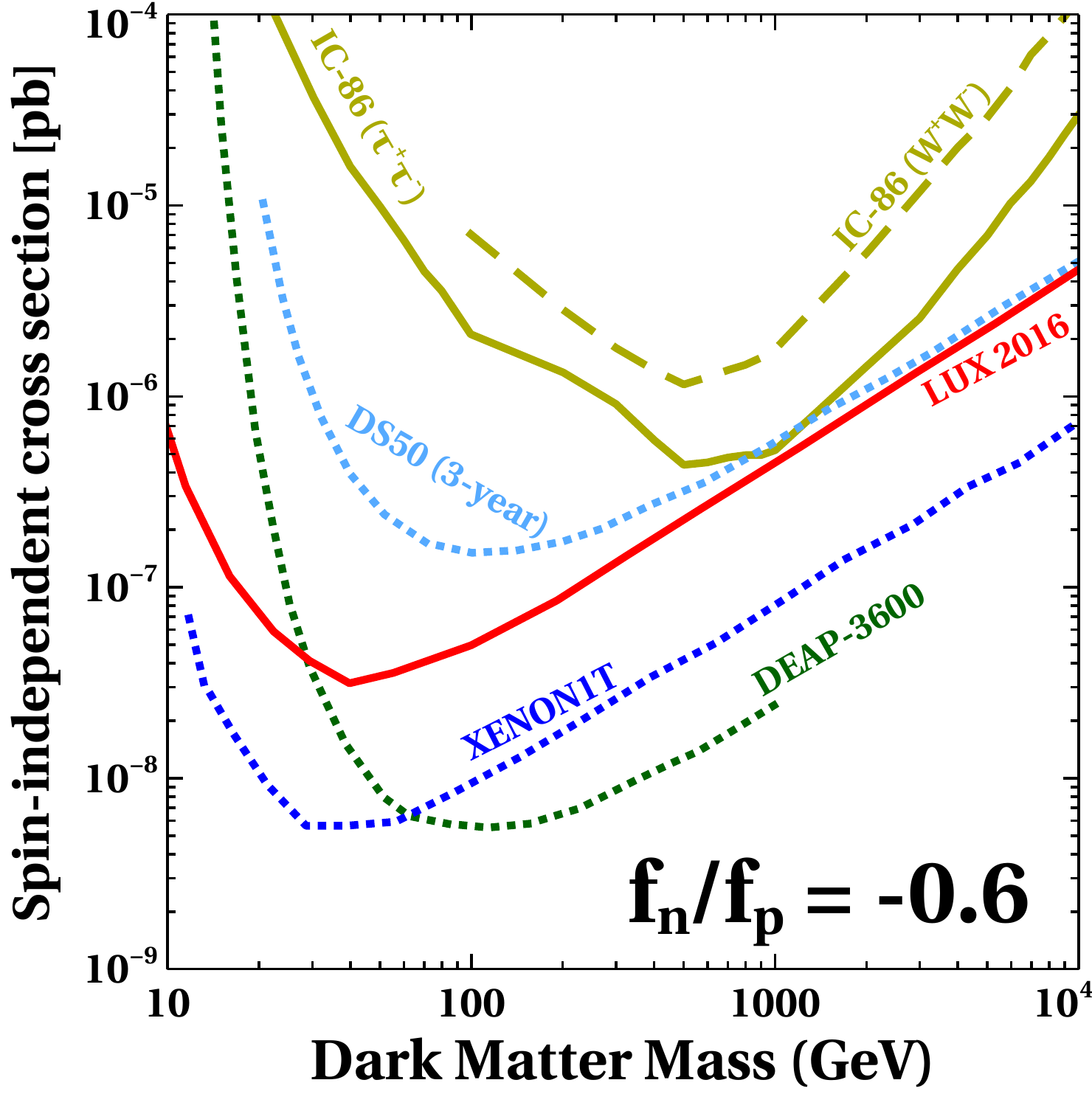} 
\caption{\label{fig:newfnfp06}  A comparison between direct detection and IceCube  bounds on the dark matter-proton spin-independent cross section for  $\ratio = -0.6$.}
\end{figure}

\begin{figure}[tb]
\centering
\includegraphics[scale=0.6]{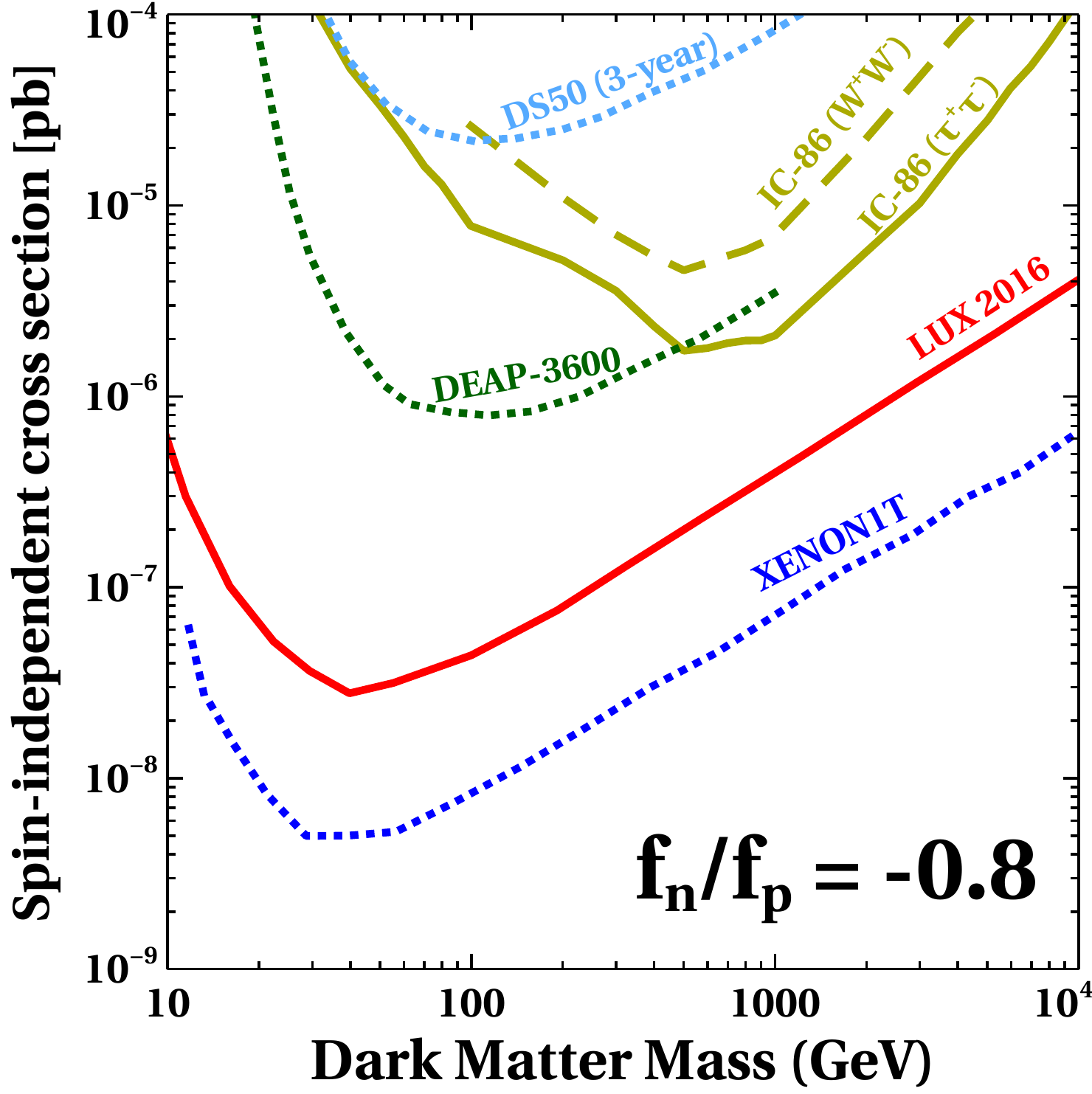}
\caption{\label{fig:newfnfp08}  A comparison between direct detection and IceCube  bounds on the dark matter-proton spin-independent cross section for  $\ratio = -0.8$.}
\end{figure}

Figures \ref{fig:newfnfp065}-\ref{fig:newfnfp08} are analogous to figure \ref{fig:fnfp07B} but for other values of $\ratio$. Figure \ref{fig:newfnfp065} shows current constraints and future sensitivities for $\ratio=-0.65$. Notice that also  in this case the  IceCube limits can be more stringent than the LUX ones, but only for dark matter annihilations into the $\tau^+\tau^-$ final state. In the future, the current LUX bound could be improved by the DarkSide-50 experiment by about a factor 2, or by the  DEAP-3600 experiment by more than one order of magnitude. Thus, also for $\ratio=-0.65$ it is the DEAP-3600 experiment, rather than XENON1T, that has the best prospects for dark matter detection.

For $\ratio=-0.75$ (see figure \ref{fig:newfnfp075}), the LUX limits become more stringent than the IceCube ones practically over the entire parameter space --the exception being at a dark matter mass of about 1 TeV and for annihilation into $\tau^+\tau^-$, where they become comparable. Notice that the 3-year sensitivity of DarkSide-50 lies entirely within the region excluded by LUX and is also partially excluded by IceCube for both annihilation channels. XENON1T and DEAP-3600 will both improve significantly the current LUX bound, reaching similar sensitivities for dark matter masses above 300 GeV.  

If $\ratio=-0.6$ the LUX limit is more stringent than the IceCube bounds and it is comparable to the 3-year expected sensitivity of DarkSide-50 at high dark matter masses, as illustrated in figure \ref{fig:newfnfp06}.  In the near future, XENON1T and DEAP-3600 will both probe new regions of the parameter space, with the latter reaching better sensitivity for dark matter masses larger than 60 GeV.

Finally, in figure \ref{fig:newfnfp08}, the results for  $\ratio=-0.8$  are displayed.  For this value of $\ratio$ the sensitivity of Argon experiments is strongly degraded --see figure \ref{fig:ddxenon}. As a result, even the projected sensitivity of DEAP-3600, let alone that of DarkSide-50,  lies deep inside the exclusion region of LUX. XENON1T is the only experiment that could discover dark matter in the near future,  probing cross sections down to $5\times 10^{-9}$ pb.

As we have seen,  in scenarios of isospin-violating dark matter, the interpretation of the experimental limits (and projected sensitivities) on the spin-independent cross section may substantially differ from those of the conventional setup. Depending on the value of $\ratio$, the hierarchy of direct detection limits obtained from different targets changes, and neutrino detectors can become more sensitive than direct detection experiments. In addition, the regions that can be probed by currently running experiments also vary with $\ratio$. All these results demonstrate, once again, that a complementary approach among  different targets in direct  detection experiments,  and between direct detection searches  and neutrino searches may be needed to unravel the nature of  the dark matter particle.

\section{Conclusion \label{sec:con}}
We presented an updated analysis of the status of isospin-violating dark matter and of  its detection prospects for the near future. Specifically,  the several experimental limits on the spin-independent cross section obtained in the last two years --from CDMS II, SuperKamiokande, DarkSide-50, PandaX, ANTARES, XENON100, IceCube and LUX-- as well as the projected sensitivity of currently operating direct detection experiments --DarkSide-50, DEAP-3600, and XENON1T-- were incorporated into the analysis. After reviewing how the interpretation of these limits and prospects is modified in isospin-violating scenarios, we compared the sensitivity of different detection experiments among themselves and against the reach of neutrino experiments.  In contrast to the standard scenario, we found, for instance, that there are regions of the parameter space where  IceCube  currently provides the most stringent constraints on the dark matter-proton spin-independent cross section, or others where the expected sensitivity of DEAP-3600 is well above the LUX exclusion limit.  Our main results are summarized in figures \ref{fig:newfnfp07}-\ref{fig:newfnfp08}. They  highlight the complementarity among  different targets in direct  detection experiments,  and between direct detection experiments  and neutrino telescopes in the search for the dark matter. 

\section*{Acknowledgments}
I would like to thank F. Queiroz for comments and suggestions. I am   supported by  the Max Planck Society in the project MANITOP.

\bibliographystyle{hunsrt}
\bibliography{darkmatter}

\end{document}